\newcommand{\ud}{{\rm d}}
\newcommand{\un}{~\mathrm}

\newcommand{\ie}{\textit{i.e.} }
\newcommand{\eg}{\textit{e.g.} }

\documentclass{elsart4-1}

 \usepackage{graphicx}

\usepackage[english,francais]{babel}
\usepackage[T1]{fontenc}
\usepackage{color}

\usepackage{amssymb}
\usepackage{amsmath} 

\renewcommand{\theequation}{\arabic{equation}}
\def\og{\leavevmode\raise.3ex\hbox{$\scriptscriptstyle\langle\!\langle$~}}
\def\fg{\leavevmode\raise.3ex\hbox{~$\!\scriptscriptstyle\,\rangle\!\rangle$}}

\begin{document}

\centerline{Physics or Astrophysics/Header}
\begin{frontmatter}


\selectlanguage{english}
\title{Dynamics of cracks in disordered materials}


\selectlanguage{english}
\author{Daniel Bonamy},
\ead{daniel.bonamy@cea.fr}

\address{SPEC, CEA, CNRS, Universit{\'e} Paris-Saclay, CEA Saclay 91191 Gif-sur-Yvette Cedex, France}


\medskip
\begin{center}
{\small Received *****; accepted after revision +++++}
\end{center}

\begin{abstract}
Predicting when rupture occurs or cracks progress is a major challenge in numerous fields of industrial, societal and geophysical importance. It remains largely unsolved: Stress enhancement at cracks and defects, indeed, makes the macroscale dynamics extremely sensitive to the microscale material disorder. This results in giant statistical fluctuations and non-trivial behaviors upon upscaling difficult to assess via the continuum approaches of engineering. 

These issues are examined here. We will see:
\begin{itemize}
\item How linear elastic fracture mechanics sidetracks the difficulty by reducing the problem to that of the propagation of a single crack in an effective material free of defects;
\item How slow cracks sometimes display jerky dynamics, with sudden violent events incompatible with the previous approach, and how some paradigms of statistical physics can explain it;
\item How abnormally fast cracks sometimes emerge due to the formation of microcracks at very small scales.
\end{itemize}
{\it To cite this article: D. Bonamy, C. R.
Physique 6 (2005).}

\vskip 0.5\baselineskip

\selectlanguage{francais}
\noindent{\bf R\'esum\'e}
\vskip 0.5\baselineskip
\noindent
{\bf Dynamique des fissures dans les matériaux désordonnés}
Prévoir quand les matériaux cassent constitue un enjeu majeur dans de nombreux domaines industriels, géologiques et sociétaux. Cela reste une question largement ouverte : la concentration des contraintes par les fissures et défauts rend en effet la dynamique de rupture à l'échelle macroscopique très sensible au désordre de microstructure à des échelles très fines. Cela se traduit par des fluctuations statistiques importantes et des comportements sous homogénéisation non triviaux, difficiles à décrire dans le cadre des approches continues de l'ingénierie mécanique.

Nous examinons ici ces questions. Nous verrons :
\begin{itemize}
\item comment la mécanique linéaire élastique de la rupture contourne la difficulté en ramenant le problème à la déstabilisation d'une fissure unique dans un matériau effectif "moyen" sans défauts; 
\item comment la fissuration lente présente dans certains cas une dynamique saccadée, composée d'événements violents et intermittents, incompatible avec l'approche précédente, mais qui peut s'expliquer par certains paradigmes issus de la physique statistique;
\item comment des fissures anormalement rapides émergent parfois du fait de la formation de microfissures à très petites échelles.
\end{itemize}
{\it Pour citer cet article~: D. Bonamy, C. R. Physique 6 (2005).}

 \vskip 0.5\baselineskip
\noindent{\small{\it Mots-cl\'es~:} fracture; solides désordonnés; crackling; lois d'échelle; transition dynamique; instabilités; approche stochastique}}
\end{abstract}
\end{frontmatter}

\selectlanguage{english}
\section{Introduction}\label{intro}

Under loading, brittle materials\footnote{In contrast with brittle fracture, ductile fracture is preceded by significant plastic deformation. Ductile fracture is always preferred in structural engineering since it involves warning. Note that the brittle or ductile nature of the fracture is not an intrinsic material property. Among others, it depends on temperature: All materials break in a brittle manner when the temperature is smaller than their so-called ductile-to-brittle transition temperature. Many catastrophic failures observed throughout history have resulted from an unforeseen crossing of this transition temperature. The sinking of the Titanic, for instance, was primarily caused by the fact the steel of the ship hull had been made brittle in contact with the icy water of the Atlantic. The loading rate is also an important parameter: Rocks behaves as brittle materials under usual conditions, but deform in a ductile manner when the loading rate becomes very small. This is \eg observed in the convection of the Earth's mantle at the origin of the plate tectonic.} like glasses, ceramics or rocks break without warning, without prior plastic deformation: Their fracture is difficult to anticipate. Moreover, stress enhancement at defects makes the behavior observed at the macroscopic scale extremely dependent on the presence of material heterogeneities down to very small scales. This results in large specimen-to-specimen variations in strength and complex intermittent dynamics for damage difficult to assess in practice. 

Engineering sidetracks the difficulty by reducing the problem to the destabilization and subsequent growth of a dominant pre-existing crack. Strength statistics and its size dependence are captured by the Weibull's weakest-link theory \cite{Weibull39_prsier} and Linear Elastic Fracture Mechanics (LEFM) relates the crack behavior to few material constants (elastic moduli, fracture energy and fracture toughness). This continuum theory provides powerful tools to describe crack propagation as long as the material microstructure is homogeneous enough and the crack speed is small enough. Conversely, it fails to capture some of the features observed when one or the other of these conditions stop being true. In particular:
\begin{itemize}
\item Slowly fracturing solids sometimes display a so-called crackling dynamics: Upon slowly varying external loading, fracture occurs by intermittent random events spanning a broad range of sizes (several orders of magnitude).
\item In the fast fracturing regime, so-called dynamic fracture regime, the limiting speed is different from that predicted by LEFM theory (see \cite{Ravichandar98_ijf,Fineberg99_pr} for reviews).  
\end{itemize}
These issues are discussed here. Section \ref{Sec1} will provide a brief introduction to standard LEFM theory, the different stages of its construction, its predictions in term of crack dynamics, the underlying hypothesis and their limitations. Crackling dynamics in slow cracks will be examined in section \ref{Sec2}. Experimental and field observations reported in this context evidence some generic scale-free statistical features incompatible with the continuum engineering approach (section \ref{Sec2.1}). Conversely, it will be seen how the paradigm of the depinning elastic interface developed in non-linear physics can be adapted to the problem (section \ref{Sec2.2}). This framework has e.g. permitted to unravel the conditions required to observe crackling in fracture (section \ref{Sec2.3}). Section \ref{Sec3} will focus on dynamic fracture and the various mechanisms at play in the selection of the crack speed. Will be seen in particular that, above a critical velocity, microcracks form ahead of the propagating crack (section \ref{Sec3.1}), making the apparent fracture speed at the macroscale much larger than the true speed of the front propagation (section \ref{Sec3.2}). It will also be seen that, at even larger velocity, the crack front undergoes a series of repetitive short-lived microscopic branching (microbranching) events, making LEFM theory not applicable anymore (section \ref{Sec3.3}). Finally, the current challenges and possible perspectives will be outlined in section \ref{concl}.  
 
\section{Continuum fracture mechanics in a nutshell}\label{Sec1}

\subsection{Atomistic point of view}\label{Sec1.1}

\begin{figure}
\begin{center}
\includegraphics[width=\textwidth]{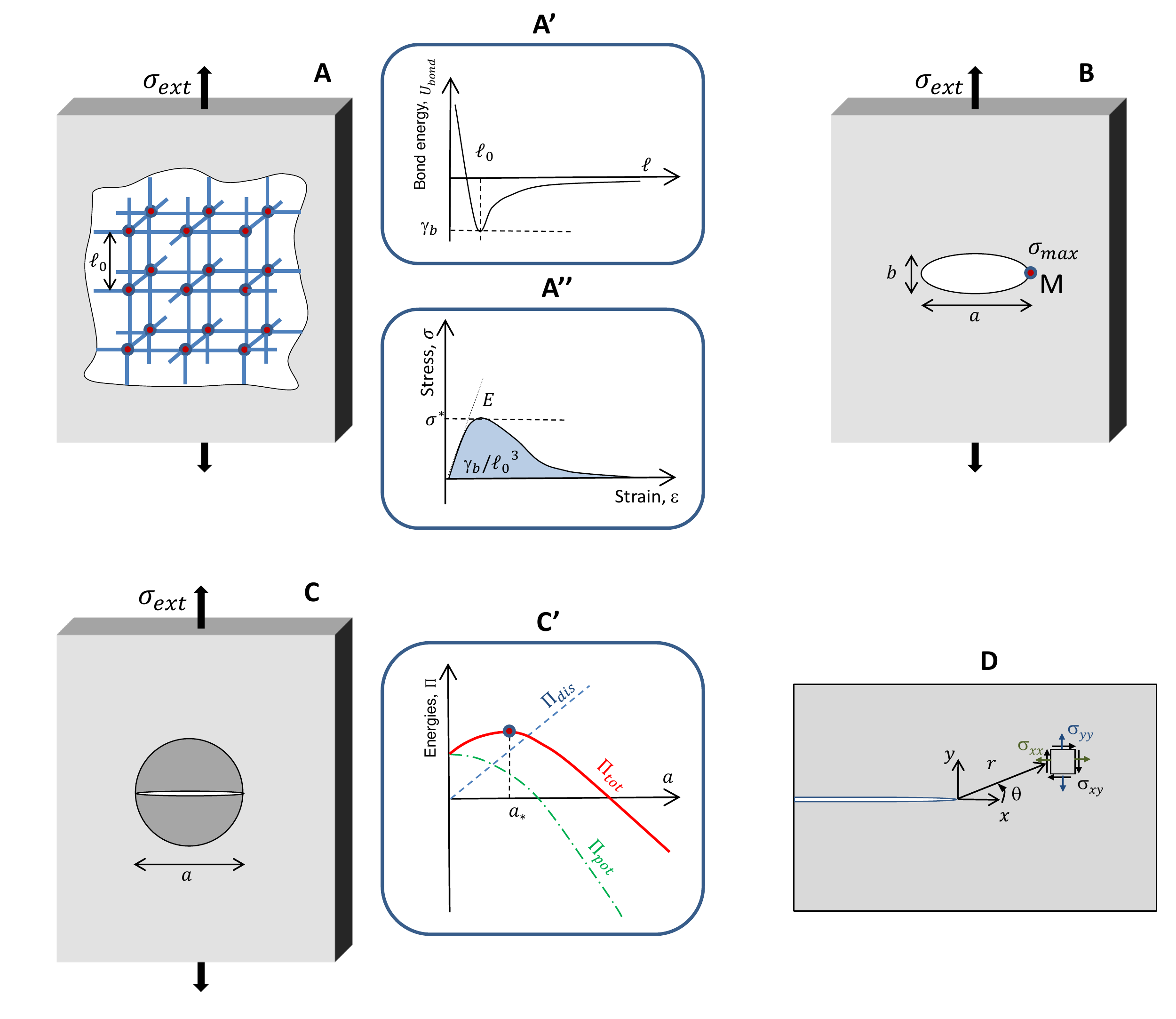}
\caption{The different stages underlying the advent of continuum fracture mechanics. Panel A: Crude atomistic view of a flawless solid loaded by a constant tensile stress, $\sigma_{ext}$. The atoms are placed on a square lattice with an interatomic distance $\ell$. They are connected by bonds whose energy, $U_{bond}$, varies with $\ell$ as depicted in panel A'. $\ell_0$ denotes the inter-atomic distance at rest, and $\gamma_b$ the associated bond energy. The "atomistic" stress-strain curve presented in panel A'' can be deduced. The slope at the origin gives the Young modulus, $E$, the area below is $\gamma_b/\ell_0^3$, and the maximum defines the strength, $\sigma_*$. Panel B: Continuum-level scale view of the solid, which now includes in its center an elliptical hole of semi-minor axis $b$ (along the loading direction) and of semi-major axis $a$ (perpendicular to loading). As shown by Inglis, the tensile stress is maximum at the apex (point M) and given by $\sigma_{max}=\sigma_{ext}\times (1+2a/b)$.  Panel C:  Griffith's view of the crack problem: The semi-minor axis $b$ goes to zero so that the elliptic defect reduces to a slit crack of length $a$. Its presence in the stressed plate leads to the release of the stress in a roughly circular zone centered on the crack with a diameter close to $a$ (dark gray zone). Panel C': The onset of crack growth is given from the comparison between two energies: The potential energy $\Pi_{pot}$ (dash green) decaying as $a^2$ (\ie as the area of the released zone) and the energy to create new fracture surfaces $\Pi_{surf}$ (dot blue), increasing linearly with $a$. For small $a$, the total energy, $\Pi_{tot}=\Pi_{pot}+\Pi_{surf}$ increases with $a$ and the crack does not move. For large $a$, $\Pi_{tot}$ decreases with increasing $a$ and the crack extends. The motion onset is at $a_*$, so that $\ud\Pi_{tot}/\ud a(a_*)=\ud \Pi_{pot}/\ud a(a_*)+\ud \Pi_{surf}/\ud a(a_*)=0$. Panel D: Notations used to describe the stress field near the tip of a slit crack.}
\label{fig1}
\end{center}
\end{figure}

Here is how strength would be inferred in a perfect solid. Take a plate pulled by an external uniform stress $\sigma_{ext}$. This plate is made of atoms connected by bonds (Figure \ref{fig1}A). As depicted in figure \ref{fig1}A' the bond energy, $U_{bond}$, evolves with the interatomic distance, $\ell$, so that the curve presents a minimum, $\gamma_b$, at a given value, $\ell_0$. This $\ell_0$ gives the interatomic distance at rest. To estimate the way the plate deforms under $\sigma_{ext}$, recall that stress is a force {\em per} surface and, hence, relates to the pulling force $F_{bond}$ via $\sigma_{ext} = F_{bond}/\ell_0^2$. Recall also that strain, $\epsilon$, is a relative deformation and, as such, relates to $\ell$ via $\epsilon=(\ell-\ell_0)/\ell_0$. Recall finally that $U_{bond}$ is a potential energy (analog to the potential energy of a spring) and, as such, relates to $F_{bond}$ via $F_{bond}=-\ud U_{bond}/\ud \ell$. The so-obtained stress-strain curve is represented in figure \ref{fig1}A''. By definition, its maximum is the sought-after strength, $\sigma_*$. 

Recall now that the {\em Young's modulus}, $E$, is the slope at origin of the curve $\sigma\,vs.\,\epsilon$, and note that, by construction, the integral $\int_0^\infty \sigma(\epsilon)\ud\epsilon$ is equal to $\gamma_b/\ell_0^3$. Introduce here the {\em free surface energy}, which is the energy to pay (in bond breaking) to create a surface of unit area:  $2\gamma_s = \gamma_b/\ell_0^2$ (the factor two, here, comes from the fact that breaking one bond creates two surface atoms). As a crude approximation and to allow analytical computation, approximate now $\sigma(\epsilon)$ by a sine: $\sigma(\epsilon) \approx \sigma^*\sin(2\pi\epsilon/\lambda)$ over the interval $0 \leq \epsilon \leq \lambda/2$: The relation $\ud \sigma /\ud \epsilon (\epsilon = 0) = E$ imposes $\lambda = 2 \pi \sigma_*/E$. The value $\sigma_*$ to make $\int_0^\infty \sigma(\epsilon)\ud\epsilon = 2\gamma_s/\ell_0$ is:

\begin{equation}
\sigma_* \approx \sqrt{\frac{E \gamma_s}{\ell_0}}
\label{Sec1:equ1}
\end{equation}        

Consider now soda-lime glass as a simple, representative example of brittle materials. The Young's modulus is about $70\un{GPa}$ and the surface energy is about $0.1\un{J/m}^2$. Taking a typical interatomic distance of $1\textup{\AA}$ leads to a theoretical strength $\sigma_* \approx 8\un{GPa}$. This value is two orders of magnitudes larger than the practical strength of the material, $\sim 50\un{MPa}$. This discrepancy is observed in almost all brittle solids! 

The element missed in the above analysis is the effect of flaws, which can make the local stress much higher than $\sigma_{ext}$. G. E. Inglis was the first, in 1913, to address this effect \cite{Inglis13_tina}: He introduced an elliptical hole in the middle of the pulled plate (Figure \ref{fig1}B) and found that the stress is maximum at the narrow ends (point M in figure \ref{fig1}B): $\sigma_{max} = \sigma_{ext}(1+2a/b)$ where $b$ and $a$ are the semi-minor (along the loading direction) and semi-major axis of the ellipse. If now, the ellipse is turned into a flaw as A.A. Griffith did in 1920 \cite{Griffith20_ptrs}, the amplification factor $a/b$ becomes tremendous. It is interesting here to introduce the radius of curvature which, for an ellipse, is $\rho=b^2/a$ at M. $\sigma_{max}$ now writes $\sigma_{max} \approx 2\sigma_{ext} \sqrt{a/\rho}$. Hence, the effect of a Griffith flaw of length $a$ and a radius of curvature of atomic dimension $\ell_0$ is to turn equation \ref{Sec1:equ1} to: 

\begin{equation}
\sigma_* \approx \frac{1}{2}\sqrt{\frac{E \gamma_s}{a}}
\label{Sec1:equ2}
\end{equation}

Flaws of micrometric size allow explaining the practical strength measured in glass.           

\subsection{Energy approach and Griffith theory}\label{Sec1.2}

The above analysis underlies the importance of flaws in determining the material strength. As such, this is not the best quantity to look at to assess material failure. Griffith hence proposed to reduce the problem of how materials fail to that of how a preexisting crack extends in a material. He addressed this question by looking at the total system energy, $\Pi_{tot}$, and how it evolves with the crack length, $a$. Two contributions are involved: 

\begin{itemize}
\item[(1)] The potential energy $\Pi_{pot}$, \ie the elastic energy stored in the pulled plate.    
\item[(2)] The energy to pay, $\Pi_{surf}$, to create the two crack surfaces.    
\end{itemize}

\noindent Their typical evolution is sketched in figure \ref{fig1}C'. The crack makes the stress release in a roughly circular zone of diameter $a$ (gray disk in Figure \ref{fig1}C): $\Pi_{pot}$ decreases  as $-a^2$. The energy cost per unit of fracture surface is, by definition, the free surface energy $\gamma_s$. Hence, $\Pi_{surf} = 2 \gamma_s a  L$ ($L$ being the plate thickness, introduced to respect the natural units of $\Pi_{surf}$ and $\gamma_s$, in J and J/m$^2$, respectively). Hence, $\Pi_{tot}$ increases with increasing $a$ when $a$ is small enough, below a critical value $a_*$ and a small crack will remain stable. Conversely, $\Pi_{tot}$ decreases with increasing $a$ for $a \geq a_*$ and a large enough initial flaw will naturally extend under the applied stress $\sigma_{ext}$. The critical value $a_*$ is the position of the maximum, so that $\ud \Pi_{tot} /\ud a|_{a_*} = \ud \Pi_{pot} /\ud a|_{a_*} + \ud \Pi_{surf} /\ud a|_{a_*} = 0$. Griffith then introduced the {\em energy release rate}, $G$, which is the amount of potential energy released as the crack advances over a unit length: 

\begin{equation}
G = -\frac{1}{L}\frac{\ud \Pi_{pot}}{\ud a}
\label{Sec1:equ3}
\end{equation}

\noindent Then, Griffith's energy criterion for crack initiation writes:

\begin{equation}
G \geq 2\gamma_s, 
\label{Sec1:equ4}
\end{equation}

\noindent and the critical size $a_*$ coincides with $G(\sigma_{ext},a_*)=2\gamma_s$. 

The next step is to determine $G$ and its evolution with $a$. In general, this is a very difficult problem to tackle analytically. However, it can be done in the situation depicted in figure \ref{fig1}C when the plate dimensions are very large with respect to $a$. In this case and in the absence of a crack, the stress is roughly identical everywhere, equal to $\sigma_{ext}$. The density of elastic energy is then $\sim \sigma_{ext}^2/E$ everywhere. As it was seen above, the introduction of the crack makes the stress release in a circular zone of diameter $a$. Hence, $\Pi_{tot}$ decreases as  $- \pi a^2 L \sigma_{ext}^2/4E$,  and equation \ref{Sec1:equ3} gives $G \approx \pi a \sigma_{ext}^2 /2 E$. Griffith's criterion allows relating $a$, $\gamma_s$ and the strength $\sigma_*$ which is the value $\sigma_{ext}$ at the point where $G = 2\gamma_s$. It gives:

\begin{equation}
\sigma_* \approx \sqrt{\frac{4 E \gamma_s}{\pi a}}, 
\label{Sec1:equ5}
\end{equation}

\noindent which is consistent with the atomistic description (equation \ref{Sec1:equ2}). The difference by a factor $4/\sqrt{\pi}$ results from the crude assumptions made both in the atomistic description and in the computation of $G$. 

\subsection{Linear elastic crack-tip field, stress intensity factor and equation of motion}\label{Sec1.3}

The main limitation of the above theory is the difficulty to determine $G$ in practical situations, when the structure exhibits a complicated geometry and/or complex loading conditions. The analysis of the stress field near the crack tip provides more powerful methods in this context. This task was first carried out by G. R. Irwin (1957) \cite{Irwin57_jam}. He considered a slit crack embedded in a 2D isotropic linear elastic solid loaded in tension (figure \ref{fig1}D) and found that the stress field exhibits a mathematical singularity
\footnote{To derive equation \ref{Sec1:equ6}, one has to find the stress field solutions  which:
\begin{itemize}
\item Obey the equations of isotropic linear elasticity, namely the equilibrium equations for stress, the compatibility equation for strain and Hook laws relating stress and strain;
\item Are compatible with the boundary conditions imposed by the crack: $\sigma_{yy}(r,\pm \pi)=\sigma_{xy}(r,\pm \pi)=0$. 
\item Is symmetric under reflection about the $x$-axis as imposed by the tensile loading along $y$: $\sigma_{xx}(-x,y)=\sigma_{xx}(x,y)$, $\sigma_{yy}(-x,y)=\sigma_{yy}(x,y)$ and $\sigma_{xy}(-x,y)=-\sigma_{xy}(x,y)$.
\end{itemize}
\noindent A possible way to do it is to introduce the Airy stress function $\Phi(x,y)$ such that $\sigma_{xx}=\partial^2_{yy}\Phi$, $\sigma_{yy}=\partial^2_{xx}\Phi$, $\sigma_{xy}=-\partial^2_{xy}\Phi$. Equilibrium equations are then satisfied automatically and the compatibility equation takes the form of a biharmonic equation: $\nabla^4 \Phi = 0$. Look for solutions of the form $\Phi(r,\theta)=r^\lambda f(\theta,\lambda)$. The boundary conditions lead to $\lambda=n/2+1$ where $n$ is an integer. Supplemented with the symmetry principles, they also provide the corresponding functions $f(\theta,n)$. As a consequence, stress and strain scale with $r$ as $\sigma_{ij} \sim \epsilon_{ij} \sim \partial^2_r \Phi \sim r^{n/2-1}$ and displacement scales as $u_i \sim r^{n/2}$. The first allowed term is that with $n=1$ since it is the first term so that $u_i$ vanish for $r\rightarrow 0$. This results in $\sigma_{ij} \sim 1/\sqrt{r}$. The functions $F_{ij}(\theta)$ in equation \ref{Sec1:equ6} are then deduced from the knowledge of $f(\theta,n=1/2)$.}: 
 
\begin{equation}
\sigma_{ij}(r,\theta) \underset{r \rightarrow 0}{\sim} \frac{K}{\sqrt{2 \pi r}} F_{ij}(\theta),
\label{Sec1:equ6}
\end{equation}

\noindent where $(r,\theta)$ are the polar coordinates in the frame $(\vec{e}_x,\vec{e}_y)$ centered at the crack tip. Here, the basis is chosen so that $\vec{e}_x$ is parallel to the direction of crack growth and $\vec{e}_y$ is parallel to the direction of applied tension. The functions $F_{ij}(\theta)$ are generic, they depend neither on the specimen geometry, nor on the crack length, nor on the elastic moduli; their form can e.g. be found in the chapter 2 of Lawn's textbook \cite{Lawn93_book}. Conversely, the prefactor $K$, so-called {\em stress intensity factor}, depends on both applied loading and specimen geometry. This is the relevant quantity characterizing the prying force acting on the crack. It relates to the energy release rate by
\footnote{Irwin proposed \cite{Irwin57_jam} the following argument to provide relation \ref{Sec1:equ7}. A crack of length $a$ can be seen as a crack of length $a+\delta a$ which is being pinched over $-\delta a \leq x \leq  0$ by applying the appropriate tractions $t(x)$. The potential energy released as the crack grows of $\delta a$ is then given by the work done by these tractions as they progressively relax to zero. Call $\Delta u_y(x)$ the opening of the crack due to $t(x)$. Equation \ref{Sec1:equ6} (supplemented with $F_{yy}(\theta=0)=1$) leads to $t(x)=K/\sqrt{2\pi(\delta a + x)}$ and equation \ref{Sec1:equ6} combined with Hook laws leads to $\Delta u_y(x) = (2 K/ E)\sqrt{-2x/\pi}$. Then, the total work done as the tractions are relaxed writes $G\delta a = \int_{-\delta a}^0 t(x) \Delta u_{y}(x) \ud x$ and equation \ref{Sec1:equ7} follows.}:

\begin{equation}
G = \frac{K^2}{E}
\label{Sec1:equ7}
\end{equation}

Returning to the plate pulled by a constant applied stress $\sigma_{ext}$ considered in the previous section, the knowledge of $G$ implies that of $K$: $K \approx \sigma_{ext} \sqrt{\pi a}$. In a more general manner, $K$ takes the form $K=\sigma_{ext} \sqrt{\pi a} \times f(a/L_i,L_j/L_i)$ where $L_i$ are the various (macroscopic) lengths involved in the system geometry: Specimen dimensions, the positions of the crack and of the loading zones, the lengths to be associated with the eventual variation of $\sigma_{ext}$ with space...  

Note that equation \ref{Sec1:equ6} implies an infinite value for the stress at the crack tip. {\em A minima}, the equation stops being valid when $r$ approaches the atomic scale $\ell_0$ since the continuum description breaks down there. In practice, in many materials, there exists a (larger) distance where, due to the singularity, stresses become so high that the material stops being elastic. The zone where this occurs is referred to as the {\em fracture process zone (FPZ)}. It embeds all the dissipative processes (plastic deformations, damage, crazing, breaking of chemical bonds, etc). Calling $\Gamma$ the total energy dissipated in this FPZ as the crack propagates so that an additional unit of surface area is created, the Griffith criterion can be generalized to:

\begin{equation}
G \geq \Gamma,
\label{Sec1:equ8}
\end{equation}

\noindent where $\Gamma$ is called the {\em fracture energy}. Note that $\Gamma$ includes the free surface energy, but can be much larger. In brittle polymers for instance where most of the dissipation comes from the disentangling of the polymer chains, $\Gamma \sim 100-1000 J/m^2$, to be compared with the typical values $\gamma_s \sim 1-10\un{J/m}^2$ for the free surface energy. Assuming as does LEFM theory that the FPZ zone is small with respect to the characteristic macroscopic scales involved in the problem ({\em small scale yielding hypothesis}), $\Gamma$ remains a material constant, independent of the specimen geometry and of the loading conditions. Since $K$ is easier to compute than $G$, the engineering community prefers to replace the Griffith's criterion by $K \geq K_c$, where $K_c=\sqrt{E\Gamma}$ defines the material toughness and, like $\Gamma$, is a material constant to be determined experimentally.

The next step is to determine the equation of motion once the crack has started to propagate. This equation is given by the balance between the total elastic energy released per unit area into the FPZ and the energy dissipated in the same zone: $G^{dyn}=\Gamma$. Note that $G^{dyn}$ includes a contribution due to kinetic energy, $\Pi_{kin}$, in addition to the potential energy considered till now.  A major hurdle here is to determine $G^{dyn}$ and the analytical solution of the associated elastodynamics problem is extremely difficult. Freund solved it in 1972 for a running crack embedded in a plate of infinite dimensions \cite{Freund72a_jmps,Freund72b_jmps,Freund73_jmps}. A first step has been to determine the near-tip stress field
\footnote{The derivation of equation \ref{Sec1:equ6dyn} is lengthly. It is \eg provided in Ravi-Chandar's book \cite{Ravichandar04_book}, in Freund's book \cite{Freund90_book} or in Fineberg and Marder's review \cite{Fineberg99_pr}.}. It exhibits a square-root singularity, similar to that in the quasi-static crack problem, which writes:  

\begin{equation}
\sigma_{ij}(r,\theta) \underset{r \rightarrow 0}{\sim} \frac{K^{dyn}}{\sqrt{2 \pi r}} F^{dyn}_{ij}(\theta,v),
\label{Sec1:equ6dyn}
\end{equation}

\noindent where the functions $F^{dyn}_{ij}(\theta,v)$ are non-dimensional generic functions indicating the angular variation of the stress field and its dependence on the crack speed, $v$. As in the quasi-static problem, they depend neither on the specimen geometry, nor on the crack length, nor on the elastic moduli. Conversely, they involve the dilational and shear wave speed, denoted $c_d$ and $c_s$, respectively. The prefactor $K^{dyn}$ is referred to as the \textit{dynamic} stress intensity factor. Remarkably, $K^{dyn}$ writes \cite{Freund73_jmps} $K^{dyn}=k(v) K$, where $k(v)$ is a universal function of $v$ (involving also $c_d$ and the Rayleigh wave speed, $c_R$), and $K$ is the static stress intensity factor that would have been obtained for a fixed crack of length equal to the instantaneous length in the same specimen geometry with the same applied loading.   

Knowing the stress field, it is possible to determine the total elastic energy per unit time flowing into the FPZ, $J=\ud(\Pi_{pot}+\Pi_{kin})/\ud t$, and subsequently $G^{dyn}=J/v$. After some manipulations\footnote{Detailed derivation of $J$ are \eg provided in Ravi-Chandar's book \cite{Ravichandar04_book} and in Fineberg and Marder's review \cite{Fineberg99_pr}. The different stages are summarized below. Using the summation convention for repeated indices, one gets $J=(\ud/\ud t)\int_A [\rho\dot{u}_i^2/2+\sigma_{ij}\partial_j u_i/2] \ud x \ud y=\int_A [\rho\dot{u}_i\ddot{u}_i+ \sigma_{ij} \partial_j \dot{u}_i] \ud x \ud y$, where $A$ denotes the area of the specimen within the $(x,y)$ plane. The equation of motion gives $\rho \ddot{u}_i=\partial_j\sigma_{ij}$. Introducing that into the integral and subsequently applying the divergence theorem leads to $J=\int_A \partial_j (\dot{u}_i\sigma_{ij})\ud x \ud y = \int_{\partial A} \dot{u}_i\sigma_{ij} n_j \ud s$ where $\partial A$ is the contour of the area $A$ and $n_i$ are the components of the outward normal to the contour in the direction of translation of the contour. Since we are interested in the energy flowing into the FPZ, we can make $A$ and $\partial A$ go to zero. Then, the stress field takes the $K$-dominant form given by equation \ref{Sec1:equ6dyn}. Combined with Hook laws, the form of the displacement components $u_i$ are deduced. It turns out that, provided a $K$-dominant form for $\sigma_{ij}$ and $u_i$, the contour integral giving $J$ is path independent. By choosing this path conveniently, one finds the relation between $J$ and $K^{dyn}$, and subsequently deduces equation \ref{Sec1:equ9old}.}, $G^{dyn}$ is found to write as the product of a universal function of $v$, $A(v)$, with the static energy release rate $G$ that would have been obtained for a fixed crack of length equal to the instantaneous length in the same system and loading geometry. At the end of the day, the equation of motion writes: 

\begin{equation}
A(v) G= \Gamma \quad \mathrm{with} \quad A(v)\approx \left( 1-\frac{v}{c_R}\right),
\label{Sec1:equ9old}
\end{equation}

\noindent Equation \ref{Sec1:equ9old}, derived for an infinite medium, is often considered as {\em the} equation of motion for cracks. This is correct in specimens of finite sizes as long as the elastic waves emitted at initiation do not have the possibility to reflect on the boundaries and come back to perturb the crack, but this is not generally correct. It is convenient to rewrite equation \ref{Sec1:equ9old} as an explicit equation of motion:

\begin{equation}
v \approx c_R \left( 1-\frac{\Gamma}{G}\right),
\label{Sec1:equ9}
\end{equation}

\noindent Recall here that $G$ quantifies the prying force acting on the crack tip. The procedure to determine how fast a crack propagates in a given geometry for a given loading is then the following: First, compute the quasi-static stress intensity factor and its variation with the crack length, by finite elements for instance (Recall here that, in a general manner, $K=\sigma_{ext} \sqrt{\pi a} \times f(a/L_i,L_j/L_i)$); Second, transform $K(a)$ into $G(a)$ using the Irwin relation \ref{Sec1:equ7}; Third, look for the value of the material-constant fracture energy $\Gamma$ (or equivalently fracture toughness $K_c$) for the considered material; Fourth, solve the ordinary differential equation \ref{Sec1:equ9} in which $v=\ud a/\ud t$ and the proper dependency $G(a)$ has been provided.  

For slow fracturing regime\footnote{In contrast with the dynamic fracture regime, the slow fracturing regime assumes a quasi-static process which can be addressed within the elastostatic framework. This approximation is relevant as long as the typical speeds in the problem (often set by the crack speed) is small with respect to the speed of the elastic waves. The Rayleigh wave speed provides a good order of magnitude for these wave speeds: It is smaller than both the dilatational and shear waves speed, and it is always very close to the latter. Slow fracturing regimes are then observed when $v \ll c_R$.}, the excess energy $G-\Gamma$ is small with respect to $\Gamma$. The expansion of equation \ref{Sec1:equ9} to the first order in $(G-\Gamma)/\Gamma$ leads to:

\begin{equation}
\frac{1}{\mu}v\simeq G-\Gamma,
\label{Sec1:equ10}
\end{equation}

\noindent where the effective mobility $\mu$ is given by $\mu=c_R/\Gamma$. 

\subsection{Limits of continuum fracture theory}\label{Sec1.5}

There are several consequences of the LEFM theory presented above: (i) Equation \ref{Sec1:equ10} predicts a continuous crack growth in the slow fracture regime and (ii) equation \ref{Sec1:equ9} suggests the Rayleigh wave speed to be the limiting speed for cracks in dynamic fracture regime. As we will see in the next sections, these two predictions are in apparent contradiction with several observations. Actually, these apparent discrepancies do not originate from flaws in the theory, but from its application hypothesis, which are: 

\begin{itemize}
\item LEFM considers the propagation of a single crack, in an otherwise homogeneous isotropic linear elastic material characterized by a material constant fracture energy.
\item LEFM is intrinsically a 2D theory and depicts the crack front as a straight line translating in a plane.   
\end{itemize}

\noindent It is only quite recently that a series of fracture experiments performed on model neo-Hookean materials (soft hydrogel) have permitted to check successfully, and quantitatively, the equation of motion predicted by LEFM over the full velocity range \cite{Goldman10_prl}. 

\section{Crackling dynamics in slowly fracturing solids}\label{Sec2}

\subsection{Short survey of experimental/field observations}\label{Sec2.1}

Equation \ref{Sec1:equ10} predicts that a crack pushed slowly in a material should propagate in a continuous and regular manner. This is not observed in a number of situations. As an illustrative example, Earth responds to the slow shear strains imposed by the continental drifts through series of sudden violent fracturing events, earthquakes. The distribution of the radiated energy presents the particularity to form a power-law, spanning many scales: 

\begin{equation}
P(E) \propto E^{-\beta}
\label{Sec2:equ1}
\end{equation}

\noindent The exponent $\beta$ slightly depends on the considered region or time\footnote{Note that earthquake sizes are more commonly quantified by their magnitude, which is linearly related to the logarithm of the energy \cite{Kanomori77_jgr}: $\log_{10}E=1.5M+11.8$. Equation \ref{Sec2:equ1} then takes the classical Gutenberg-Richter frequency-magnitude relation: $P(M)\propto \exp(-bM)$ where $b$ refers to the Richter-Gutenberg exponent and relates to $\beta$ via: $\beta=b/1.5+1$.} but always remains close to $\beta \simeq 1.6$. This kind of distribution is characteristic of scale-free systems. It has an important consequence: Its second moment, $\overline{E^2}=\int_0^\infty E^2 P(E)\ud E$, is infinite;  The notion of a typical "average" intensity for earthquakes is meaningless! Beyond the power-law distribution for energy, earthquakes also present a specific organization in time: The inter-event time is power-law distributed \cite{Bak02_prl}, and the events organizes into mainshock-aftershock sequences obeying a range of empirical scaling laws, the most common of which are: Omori-Utsu law \cite{Omori94_jcsiut,Utsu95_jpe}  (aftershock frequency decays algebraically with time from mainshock), productivity law \cite{Utsu71_jfshu,Helmstetter03_prl} (number of produced aftershocks increases as a power-law with the mainshock energy) and Bath law \cite{Bath65_tec} (difference in magnitude between mainshock and its largest aftershock is independent of the mainshock magnitude).   

Laboratory-scale experiments have revealed similar statistical features in the the acoustic emission going along with the fracture of different heterogeneous solids: Solid foams \cite{Deschanel09_jpd}, plaster \cite{Petri94_prl}, paper \cite{Salminen02_prl}, wood \cite{Makinen15_prl}, charcoal \cite{Ribeiro15_prl}, mesoporous silica ceramics \cite{Baro13_prl}, etc. The energy of the acoustic events and the silent time between them are power-law distributed. More recently, the analogy between seismology and fracture experiments at the lab-scale has been further deepened with the evidence of aftershock sequences obeying the standard laws of seismology \cite{Baro13_prl,Makinen15_prl,Ribeiro15_prl}. Note that the different scaling exponents involved in the problem were reported to depend on the considered material and fracture conditions \cite{Rosti09_jpd,Deschanel09_jpd,Bonamy09_jpd}.

Unfortunately, the relation between AE energy and released elastic energy remains largely unknown; reference \cite{Minozzi03_epjb} attempts to better understand this relationship by using minimal lattice networks. Moreover, most of the acoustic fracture experiments reported in the literature start with an intact specimen, and load it up to the overall breakdown. In these tests, the recorded acoustic events reflect more the microfracturing processes preceding the initiation of the macroscopic crack than the growth of this latter. This has motivated few groups to look at simpler 2D systems, closer to the LEFM assumptions: A group in Oslo has imaged the dynamics of a crack line slowly driven along a weak heterogeneous interface between two sealed transparent Plexiglas plates \cite{Maloy01_prl,Maloy06_prl}. They showed that the crack progresses via depinning events, the area of which is a power-law distribution. A group in Lyon has observed directly the slow growth of a crack line in 2D sheets of paper \cite{Santucci04_prl}. This occurs via successive crack jumps of power-law distributed length.

\begin{figure}
\begin{center}
\includegraphics[width=\textwidth]{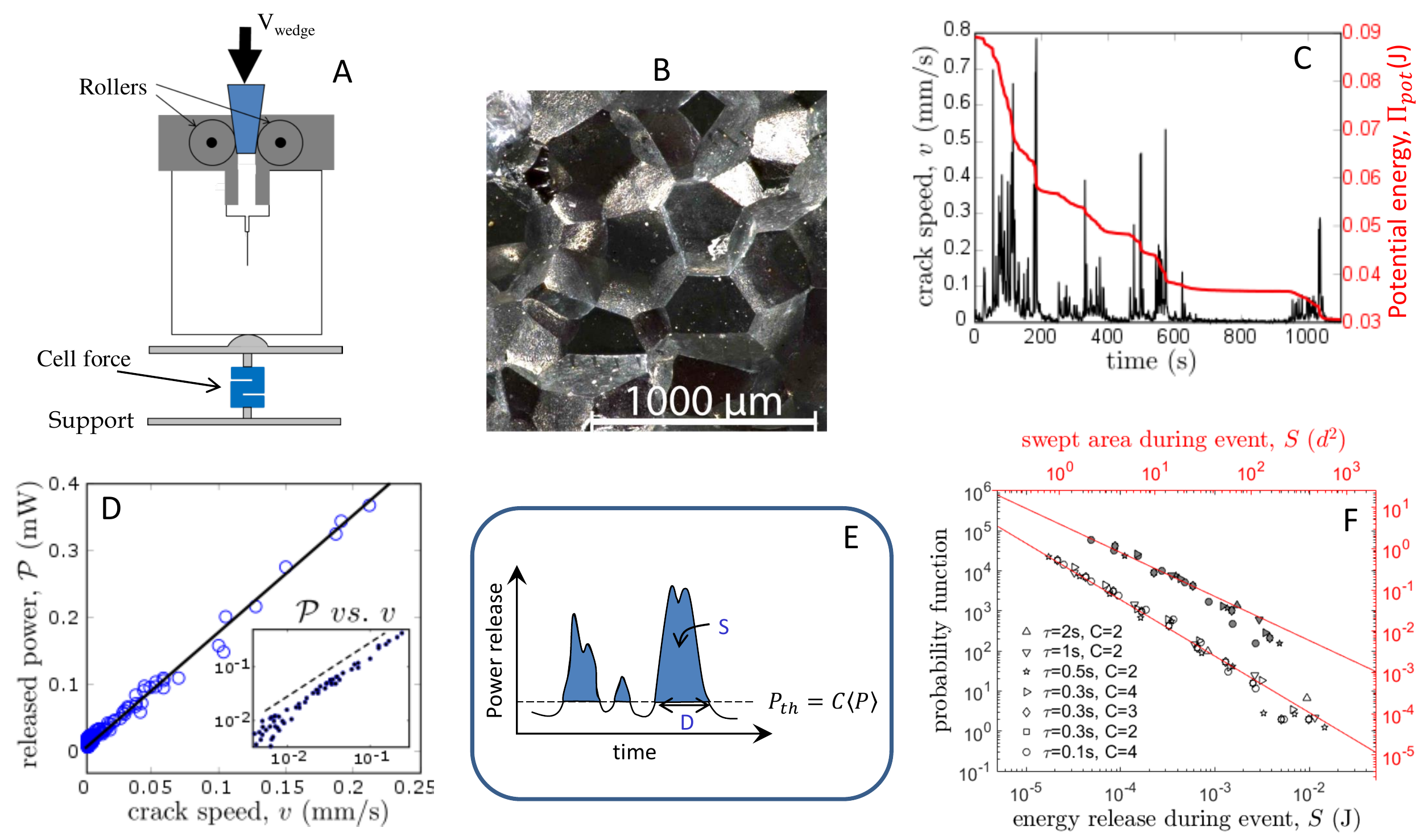}
\caption{Crackling dynamics of a slowly driven crack in an artificial rock. Panel A: Sketch of the experimental setup. Panel B:  Microscope image of the fracture surfaces. Note the facet-like structure illustrating the intergranular fracture mode and the absence of visible porosity. The diameter of the beads used to synthesis this rock was $583\,\mu\mathrm{m}$. Panel C: Zoomed view of the crack speed $v(t)$ (black) and the potential elastic energy $\Pi_{pot}(t)$ stored in the specimen (red) as a function of time in a typical fracture experiment. Panel D: Instantaneous released power $\mathcal{P}(t)=-\ud \Pi_{pot}/\ud t$ as a function of $v(t)$ for all $t$. The proportionality constant (slope of straight line) sets the fracture energy $\Gamma=100\pm 10\un{J/m}^2$. Panel E: Standard procedure to extract the avalanche size and duration from such a crackling signal within the depinning interface framework (See also figure \ref{fig3}): A threshold is prescribed and the avalanches are identified as the individual bursts above this threshold. The avalanche duration is defined from the two successive times the curve crosses this threshold. The avalanche size, $S$, is defined as the integral of the burst above the threshold.  Panel F: Distribution of $S$, expressed either as the energy released during the event (bottom x-axis) or as the area swept during the events made dimensionless by the bead diameter $d$ (top x-axis). The various symbols correspond to various coarsening times $\delta t$ and different values for the prescribed threshold $C\langle v \rangle$, where $\langle v \rangle$ is the speed averaged over the whole experiment: $\langle v \rangle=2.7\,\mu$m/s (empty symbol) and $\langle v \rangle=40\,\mu$m/s (filled symbol); the latter has been shifted vertically for sake of clarity (Adapted from \cite{Bares14_prl}).}
\label{fig2}
\end{center}
\end{figure}

More recently, our group carried out a series of crack growth experiments in artificial rocks. These rocks were obtained by mimicking the processes underlying the formation of real rocks in nature: A mold was first filled with monodisperse polystyrene beads and heated up to $105^\circ\mathrm{C}$ ($\sim 90\%$ of the temperature at glass transition). The softened beads were pressed between the jaws of an electromechanical machine at a prescribed pressure. Then, both pressure and temperature were kept constant for one hour. This gives the time for sintering to occur. The mold was then brought back to ambient conditions of temperature and pressure slowly enough to avoid residual stresses. This procedure provides artificial rocks whose microstructure length-scale and porosity are set by the bead diameter and applied pressure (see \cite{Cambonie15_pre} for details). The rock porosity was kept small enough (to a few \textit{percent}) so that fracture occurs from the propagation of a \textit{single} crack in between the sintered grains (Figure \ref{fig2}B).

In the so-obtained materials, a seed crack was initially introduced with a razor blade. This crack was slowly driven throughout the rock using the experimental arrangement depicted in figure \ref{fig2}A, by pushing at small constant speed a triangular wedge into the cut-out on one side of the sample ($10 \sim 100 \un{nm/s}$ range for the wedge speed). In this so-called wedge splitting geometry, the crack is expected to grow at a speed set by the wedge speed (about two orders of magnitude larger). In addition to the crack speed, $v(t)$, special attention was paid to monitor in real time the potential elastic energy stored in the specimen, $\Pi_{pot}(t)$. This has been made possible by placing two go-between steel blocks equipped with rollers between the wedge and the specimen. This limits parasitic dissipation via friction or plastic deformation at the contact, so that the failure processes within the FPZ are ensured to be the sole dissipation source in the system (see figure \ref{fig2}A and reference \cite{Bares14_prl} for details). 

Figure \ref{fig2}C shows the typical measured signals. They display an irregular burst-like dynamics with random sudden fluctuations spanning many scales. Again, such a highly fluctuating dynamics is incompatible with the equation \ref{Sec1:equ10} of LEFM theory. Yet and despite their individual giant fluctuations, the elastic power released $\mathcal{P}(t)=-\ud \Pi_{pot} /\ud t$, was found to be proportional to the speed fluctuation $v(t)$ at each time step (figure \ref{fig2}D). This enables defining a material-constant fracture energy $\Gamma$. As we will see in the next section, these observations can be explained within the depinning interface paradigm applied to heterogeneous fracture. To characterize the fluctuation statistics, we hence adopted the standard procedure in the field. As depicted in figure \ref{fig2}E, this consists in identifying the underlying depinning avalanches with the bursts where $\mathcal{P}(t)$ is above a prescribed reference level, $\mathcal{P}_{th}$. Then, the duration $T$ of each pulse is given by the interval between the two intersections of $\mathcal{P}(t)$  with $\mathcal{P}_{th}$, and the avalanche size $S$ is defined as the energy released during the event, i.e., the integral of $\mathcal{P}(t)$ between the two intersection points. As expected in the depinning interface paradigm (see next section), $S$ follows a power-law distribution, $P(S) \propto S^{-\tau}$ (Figure \ref{fig2}F) and $T$ scales as a power-law with $S$, $T\propto S^{\gamma}$. Let us finally mention that the acoustic emission has been also analyzed in our experiments. The acoustic events get organized to form aftershock sequences obeying the laws of seismology \cite{Bares17_submitted}.

\subsection{Depinning of elastic interfaces as a paradigm of heterogeneous fracture}\label{Sec2.2}

The experiments reported in the previous section suggest that fracturing (heterogeneous) solids belong to the so-called crackling systems. This class encompasses a variety of different systems, those which respond to slowly varying external conditions through random impulsive events of power-law distributed (scale-free) size \cite{Sethna01_nature}. This class of problems \eg includes fluctuations in the stock market\cite{Bouchaud01_qf}, paper crumpling \cite{Houle96_pre}, or cascading failure in power grids \cite{Chen05_ijepes}. 

\begin{figure}
\begin{center}
\includegraphics[width=0.7\textwidth]{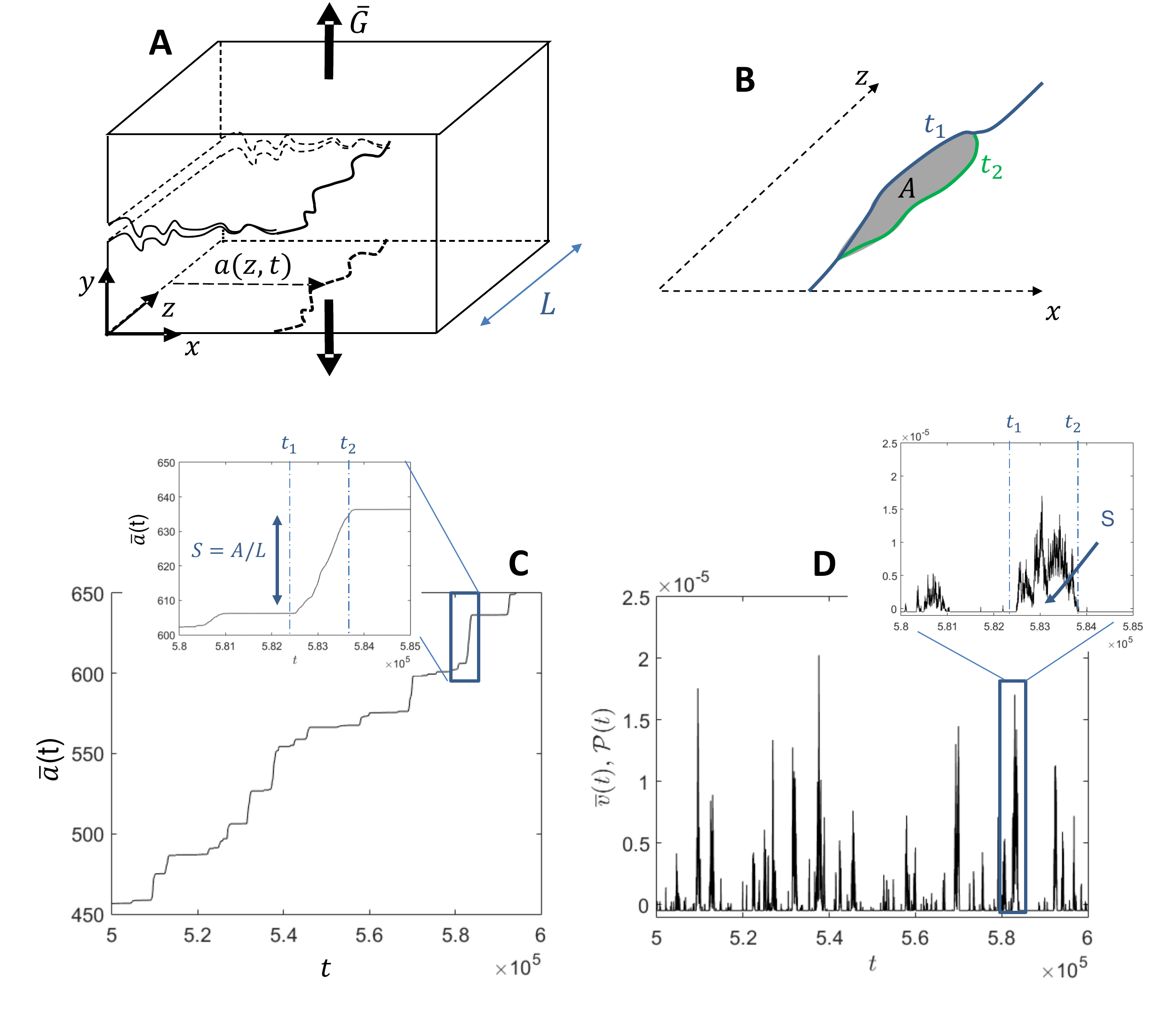}
\caption{Elastic depinning approach applied to stable crack propagation. Panel A: Sketch and notations used to derive equations \ref{Sec2:equ3} and \ref{Sec2:equ4}. The crack propagates by a series of jumps (avalanches) between successive pinned configurations. Panel B: For each avalanche, the duration, $T$, is defined by the duration of the jump and the times $t_1$ and $t_2$ coincide with the start and end of the jump. The avalanche size $S$ is set by the area $A$ swept over this jump. Panel C: As a result the time evolution of the spatially-averaged crack length $\overline{a}(t)$ exhibits a step-like form where the pinned regions coincide with the horizontal portions, and the avalanches coincide with the stiff portions. Panel D: The spatially-averaged crack speed, $\overline{v}(t)$ (resp. the instantaneous power released, $\mathcal{P}(t)$) exhibits a crackling dynamics made of successive bursts the duration of which are set by $T$. Moreover, the integral below the curve is given by $S/L$, where $L$ is the specimen thickness (resp. $S/\overline{\Gamma}$ where $\overline{\Gamma}$ is the material fracture energy) (adapted from \cite{Bonamy09_jpd}).
}
\label{fig3}
\end{center}
\end{figure}

Crackling dynamics in fracture cannot be captured by the continuum approaches of LEFM theory. We now turn to another approach -- pioneered by H.~G. Gao and J.~R. Rice in 1989 \cite{Gao89_jam} and developed firstly to explain the scale-invariant properties of crack surface roughness \cite{Bouchaud93_prl,Schmittbuhl95_prl,Larralde95_epl,Ramanathan97_prl,Bonamy06_prl,Bonamy11_pr}. The idea is to consider \textit{explicitly} the presence of inhomogeneities in the material microstructure by introducing a stochastic term, $\eta$, into the fracture energy: $\Gamma(x,y,z)=\overline{\Gamma}\times (1+\eta(x,y,z))$. The system is depicted in figure \ref{fig3}A. Note that the third dimension, $z$, is now explicitly considered, which was not the case till now. The fluctuations in fracture energy induces distortions of the front, which, in turn, generate local variations in the energy release rate, $G(z,t)$. In a first order analysis, the out-of-plane roughness of the crack line can be neglected and $G(z,t)$ depends on the in-plane component of the crack line distortions only \cite{Movchan98_ijss}. J.~R. Rice (1985) has provided the relation in the limit of a specimen with an infinite thickness \cite{Rice85_jam}: 

\begin{equation}
G(z,t)=\overline{G}(1+J(z,\{a\})) \quad \mathrm{with} \quad J(z,\{a\})=\frac{1}{\pi} PV \int_{-\infty}^{\infty} \frac{a(\xi,t)-a(z,t)}{(\xi-z)^2}\ud \xi
\label{Sec2:equ2}
\end{equation}

\noindent Here, $PV$ denotes the principal part of the integral and $a(z,t)$ is the in-plane position of the crack line (Figure \ref{fig3}A). $\overline{G}$ denotes the energy release rate that would have been used in the standard LEFM picture, after having coarse-grained the microstructure disorder and averaged the behavior along the $z$ direction, as implicitly done all along section \ref{Sec1}. In the same way and all along this section, the crack length averaged over the specimen thickness (say the standard LEFM crack length) will be referred to as $\overline{a}$. The associated LEFM-level scale crack speed will be referred to as $\overline{v}$: $\overline{v}(t)=\ud \overline{a}/\ud t$.

The application of Griffith's criterion at each point $z$ along the front provides \cite{Schmittbuhl95_prl,Ramanathan97_prl}: 

\begin{equation}
\frac{1}{\mu} \frac{\partial a}{\partial t} = F(\overline{a},t)+\overline{\Gamma} J(z,\{a\})+\overline{\Gamma}\eta(x=a(z,t),z),
\label{Sec2:equ3}
\end{equation}

\noindent where $F(\overline{a},t)=\overline{G}(\overline{a},t)-\overline{\Gamma}$. Now, look at the spatially averaged solution $\overline{a}(t)$  of this equation (or its derivative $\overline{v}(t)$). A rapid and naive glance would suggest that taking the average of this equation naturally leads to the LEFM equation of motion (equation \ref{Sec1:equ10}): The averaging of the term $a(\xi,t)-a(z,t)$ in the integral term of equation \ref{Sec2:equ2} indeed makes the second right-handed term vanish, and the mean value of $\eta$ is, by definition, zero. These arguments are not correct! The reason is that the stochastic term $\eta$ is a \textit{frozen} disorder term; it depends explicitly on the position of the crack line. Therefore, the coarse-graining of equation \ref{Sec2:equ3} should properly account for the fact this line stays pinned much longer at the strongest points in $\eta$, which, hence, counts much more in the averaging process. 

The solution of this equation is known \cite{Ertas94_pre} to exhibit the coined depinning transition governed by $F$ and its relative position with respect to a critical value $F_c$:
\begin{itemize}
\item when $F$ is smaller than $F_c$ the front is pinned by the disorder and does not propagate;
\item when $F$ is much larger than $F_c$ the front grows at a mean speed $\bar{v}$ proportional to $F$. In other words, $\overline{v}$ is proportional to $\overline{G}-\overline{\Gamma}$ and the standard LEFM theory (equation \ref{Sec1:equ10}) is recovered;
\item when $F$ is exactly equal to $F_c$, a {\em critical} state is observed, and the system becomes scale-invariant in both space and time. At this peculiar point, the crack line moves via jumps the statistics of which is power-law distributed over the full accessible range, from the scale of the microstructure to the specimen size.
\item when $F$ is larger than $F_c$, but not too much, the scale-invariant features of criticality only extend up to a finite upper cutoff. This cutoff is all the more important so as the difference $F-F_c$ is small.   
\end{itemize}
\noindent This critical state close to $F_c$ explains the crackling dynamics in fracturing solids. More importantly, it allows us to invoke the {\em universality} of the crack dynamics there. The different exponents involved in the power-law distribution of size and duration and in the scaling between the two are generic: They depend neither on the "microscopic details" of the system (say the precise size and shape of inhomogeneities, their nature...), nor on the "macroscopic details (the way the solid is loaded, for instance)
\footnote{The independence from the system details deserves some comments: It will remains true provided that: (i) There exists a well-defined spatial correlation length for the frozen disorder (or well-defined correlation lengths along the relevant directions in the case of an anistropic microstructure); (ii) this correlation length (or these correlation lengths) is small with respect to the macroscopic dimensions involved in the system; and (iii) the scaling properties are looked at scales above this correlation length (or well above the largest of these correlation lengths). It will not work in laminar materials or in the complex hierarchical biological structures like bones for instance. Recall also that equations \ref{Sec2:equ3} and \ref{Sec2:equ4} were derived in the framework of isotropic linear elasticity and the effect of the material inhomogeneities is integrated in the fracture energy only (or equivalently in the fracture toughness). Inhomogeneities in elastic properties, for instance, are not included. Recall finally that a slow fracturing regime is considered here and that the inertial effects due to the elastic waves are not considered. Those can be very important, \eg in the triggering of earthquakes.}. Even more surprisingly, these exponents are identical to those observed in other systems belonging to the same universality class, \eg the contact line motion in wetting \cite{Joanny84_jcp,Ertas94_pre} and domain wall motion in ferromagnets \cite{Urbach95_prl,Durin00_prl}. Last but not least, these exponents are predictable (to some extend) using the functional renormalization group (FRG) methods initiated by D.~S. Fisher \cite{Fisher86_prl} and further developed, \eg in \cite{Chauve01_prl,Rosso09_prb,Dobrinevski12_pre}.   

From the above analysis, a crackling dynamics is expected in fracture, provided the system remains close to the critical point. But why should it be the case? To understand this, in Ref. \cite{Bonamy08_prl}, we took a closer look at the form of the term $F(\overline{a},t)$ in equation \ref{Sec2:equ3}. Recall here that$\overline{G}$ goes approximately as $\sigma_{ext}^2\overline{a}$ (sections \ref{Sec1.2} and \ref{Sec1.3}); it is an increasing function of the crack length. This means that any system loaded by imposing a constant stress $\sigma_{ext}$ would yield an unstable fracture, with a crack accelerating very rapidly up to its limiting speed (equation \ref{Sec1:equ9} in section \ref{Sec1.3}). Slow fracturing situations, hence, can only be encountered in systems loaded by imposing a \textit{time-increasing external displacement}, $u_{ext}=v_{ext} t$. Then, the ratio $k=\sigma_{ext}/u_{ext}$, referred to as the specimen stiffness, is a decreasing function of the crack length. In slow fracturing situations, the decrease of $k$ with $\overline{a}$ overcomes the linear increase of $\overline{G}$ with $\overline{a}$ and, finally, $\overline{G}(\overline{a},v_{ext} t)$ decreases with increasing $\overline{a}$. In a first approximation, the term $F$ in equation \ref{Sec2:equ3} writes \cite{Bonamy08_prl}:

\begin{equation}
F(\overline{a}(t),t) \approx \dot{G}t-G'\overline{a},
\label{Sec2:equ4}
\end{equation}

\noindent where $\dot{G}=\partial \overline{G}/\partial t$ and $G'=-\partial \overline{G}/\partial \overline{a}$ are positive constants set by the external displacement field and the specimen geometry, only. The crack motion can then be decomposed as follow: As long as $F(\overline{a}(t),t) \leq F_c$, the front remains pinned and $F(\overline{a}(t),t)$ increases with $t$. As soon as $F \geq F_c$, the front starts propagating, making $\overline{a}$ increase and $F(\overline{a}(t),t)$ decrease. As we will see in the next section, there exists a whole range for the parameters $\dot{G}$ and $G'$ so that these two antagonist mechanisms maintain $F$ close to the critical point during the whole propagation \cite{Bonamy08_prl}. A self-sustained steady crackling dynamics made of depinning avalanches is then observed (Figure \ref{fig3}B). In this picture, the area $A$ of these avalanches sets both the jump size for $\overline{a}$: $S=A/L$ and the potential energy released during this jump: $\delta \Pi_{tot}=\overline{\Gamma} A$ (Figures \ref{fig3}C and D). This picture is consistent with the observations reported at the end of the previous section in artificial rocks \cite{Bares14_prl} where, despite their giant fluctuations, $\overline{v}(t)$ and $\mathcal{P}(t)$ were remaining proportional at all times (figure \ref{fig2}D). 

Within this framework, $S$ (or equivalently $A$ or $\delta\Pi_{pot})$ is power-law distributed: $P(S) \propto S^{-\tau}$ with $\tau = 1.280\pm0.010$ \cite{Bonamy09_jpd}. This power-law extends over the full scale range when $\dot{G} \rightarrow 0$ and $G' \rightarrow 0$. More generally, the distribution writes $P(S) \propto S^{-\tau}f(S/S_0)$ where $f(u)$ is a quickly decreasing function and $S_0 \simeq g(\dot{G}/G') G'^{-1/\sigma}$ where $g(u)$ is an increasing function and  $\sigma=  1.445\pm 0.005$ \cite{Bares14_frontiers}. Moreover, the avalanche duration $T$ scales with $S$ as $T \propto S^\gamma$ with $\gamma = 0.555\pm 0.005$ \cite{Bonamy09_jpd}. The power-law behaviors for $P(S)$ and $T\,vs\,S$ are consistent with what was observed in the fracture of the artificial rocks (Figure \ref{fig2}F and previous section). Conversely, the exponent value measured experimentally is significantly different. This discrepancy is thought to result from the finite width $L$ of the experimental fracture specimen not taken into account in the derivation of equation \ref{Sec2:equ3}. 

\subsection{From crackling to continuum-like dynamics}\label{Sec2.3}

The preceding section has permitted to show that crackling dynamics {\em may} emerge from the interactions between a crack and the material inhomogeneities. Still, actual experimental observations of crackling are scarce and most situations involving stable crack growth in a variety of disordered brittle solids (structural glasses, brittle polymers, ceramics,...) exhibit a continuous dynamics compatible with the LEFM predictions. 

\begin{figure}
\begin{center}
\includegraphics[width=\textwidth]{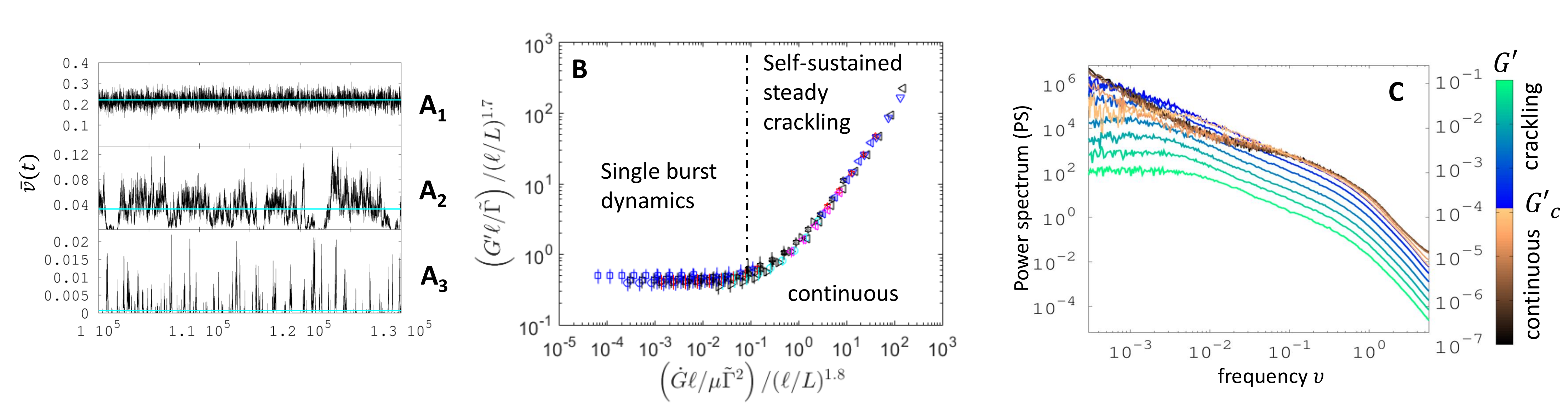}
\caption{Crackling vs. continuum dynamics in heterogeneous fracture. Panel A: Time evolution of the spatially-averaged velocity $\overline{v}(t)$ predicted by equations \ref{Sec2:equ3} and \ref{Sec2:equ4} for increasing values of $G'$: $G'=4.75\times 10^{-5}$ (A1), $G'=2\times 10^{-4}$ (A2), $G'=5.5\times 10^{-3}$ (A3). The other parameters are kept constant: $\dot{G}=10^{-5}$, $\overline{\Gamma}=1$, $\mu=1$, $L=1024$, and $\eta(x,z)$ is an uncorrelated random landscape of zero average and unit variance. At low $G'$, $\overline{v}(t)$ wanders around the value $G'/\dot{G}$, as predicted within the LEFM framework. When $G'$ increases, the dynamics becomes jerky and switches to crackling dynamics made of separate pulses the duration of which decreases with increasing $G'$. Panel B: Phase diagram of the crack dynamics predicted within the depinning interface framework (Equations \ref{Sec2:equ3} and \ref{Sec2:equ4}). This diagram is fully defined by two reduced variables mingling all the parameters involved in equations \ref{Sec2:equ3} and \ref{Sec2:equ4}. Panel C:  Fourier spectrum of $\overline{v}(t)$ at increasing $G'$ (value indicated in the right-handed colorbar), keeping all the other parameters constant (same value as in panel A1$\rightarrow $A3). Note the qualitative change as the transition line in panel B is crossed (i.e. as $G'$ crosses $G'_c$). Note also that only the lowest frequencies of the spectra evolve with $G'$ below $G'_c$. Note finally the power-law characteristic of a scale-free dynamics above $G'_c$ (Adapted from \cite{Bares13_prl}).}
\label{fig4}
\end{center}
\end{figure}

To shed light on when crackling dynamics is likely to occurs, we have numerically explored the parameter space associated with equations \ref{Sec2:equ3} and \ref{Sec2:equ4}. Crackling is favored by \cite{Bares13_prl}:
\begin{itemize}
\item Larger disorder or heterogeneities, \ie larger standard deviation or larger spatial correlation length for the random frozen landscape $\eta(x,y,z)$ in equation \ref{Sec2:equ3};
\item Smaller thickness for the fracturing specimen, or to be more precise smaller ratio thickness over heterogeneity size. The small value of this ratio ($\sim 30$) is what has permitted to observe crackling in the artificial rocks of figure \ref{fig2}. The downsizing of high-tech mechanical components is \eg anticipated to favor crackling and unpredictability against continuum dynamics and predictable fracture behavior.   
\item Fracture geometry so that $G$ decreases rapidly with crack length (smaller $G'$ in equation \ref{Sec2:equ4}, see also figure \ref{fig4}A$_1\rightarrow$A$_3$). This is e.g. achieved in indentation problems. This effect may be the one responsible of the unforeseen earthquake-like fracturing events observed at the keV$\sim$MeV scale in a cryogenic detector during the early stages of the CRESST experiment searching for dark matter in high energy physics \cite{Astrom06_pla}. 
\item Slower rate for the displacement imposed externally, \ie smaller $\dot{G}$ in equation \ref{Sec2:equ4}. This is typically the situation encountered in seismology.  
\end{itemize}  
\noindent  A careful analysis of the above simulations combined with dimension analysis has allowed us to unravel the transition line between the crackling and LEFM-like regimes. It defines a phase diagram within a space defined by two reduced variables only, represented in figure \ref{fig4}B. 

Finally, it is of interest to look at the Fourier spectrum of $\overline{v}(t)$ and how it evolves when the transition line is crossed. This is depicted in figure \ref{fig4}C, presenting a series of spectra at increasing values $G'$ (recalled to be the derivative of $G$ with crack length), \ie taken along a vertical line in the phase diagram of figure \ref{fig4}B. Here, $G'_c$ denotes the value at the transition line, the dark-to-light brown curves correspond to the spectra observed in the LEFM-like phase, below $G'_c$, and the blue-to-green curves correspond to the spectra observed in the crackling phase, above $G'_c$. In the LEFM-like phase, all curves collapse except at the lowest frequencies. This is what must be true  in  a  continuum  description where  a macroscopic control parameter (here $G'$) should affect the system at the larger scales only (small frequencies). Conversely, in the crackling phase, changing $G'$ affects \textit{all} the scales (all the frequencies) and the curves do not overlap at \textit{any} place. Note the power-law form of the spectra (straight line in logarithmic scales) characteristic of a scale-free dynamics. Note also the fact that the power-law exponent remains unaffected by the increase of $G'$, as expected from the universality invoked in the previous section. Note finally the suddenness of the changes observed as the transition line is crossed; a special attention was paid, in the plot of figure \ref{fig4}C, to modulate $G'$ in a regular manner, by increasing it by the same multiplicative constant all along the process. This suddenness on the aspect change suggests an underlying true transition rather than a simple crossover phenomenon. This LEFM-to-crackling transition is distinct from the standard depinning transition; it occurs within the depinned phase, at a finite (but small) value of the mean front speed. Future work is required to fully characterize the underlying mechanisms.    

\section{Damage-induced boosting of fast cracks}\label{Sec3}

We now turn to dynamic fracture and the effect of microstructure disorder onto the continuum-level scale dynamics, when $v$ reaches a value of the order of $c_R$ (say larger than $c_R/10$). According to the phase diagram uncovered in section \ref{Sec2.3} and figure \ref{fig4}B, the giant velocity fluctuations due to the interface depinning mechanism disappear as $v$ is sufficiently large and LEFM predictions are recovered. As we will see below, another mechanism is activated within the FPZ, bringing a new source of complexity. 

\subsection{Velocity-induced nominally-to-quasi brittle transition in disordered solids}\label{Sec3.1}

As recalled in section \ref{Sec1.3}, LEFM theory predicts the limiting crack speed, $v_\infty$, to be $c_R$. In practice, this is not observed and $v_\infty$ is often reported to range between $0.5\sim 0.6 c_R$ \cite{Ravichandar04_book}. To understand this discrepancy, we carried out a series of dynamic fracture experiments in Polymethylmethacrylate (PMMA)\footnote{PMMA is often considered as the archetype of nominally brittle materials in experimental mechanics. The size of the FPZ is very small, around $30\,\mu$m. At ambient temperature, the viscosity is very small and the mechanical behavior is well described by isotropic linear elasticity, even down to very small scales. Its elastic modulus, $E\sim 3\un{GPa}$, is large enough so that the specimens can be easily manipulated and do not deform significantly under their own weight. At the same time, $E$ is small enough so that, during testing, the specimen deformations are large enough compared with those of the different pieces that constitute the loading machine (generally in steel). Moreover, PMMA is transparent which allows the direct imaging of the processes, and birefringent which also permits to image the stress field (or more exactly the deviatoric part of the stress tensor). Last but not least, PMMA is cheap. For all these reasons, PMMA has been one of the most widely used materials against which theories have been confronted with from the early stages of fracture mechanics.}, with the aim of measuring both $K(t)$ and $v(t)$ independently and testing equation \ref{Sec1:equ9} quantitatively \cite{Scheibert10_prl}. In this context, we used the wedge-splitting experimental arrangement depicted in figure \ref{fig2}A with three adaptations due to the constraints of dynamic fracture and the requirement of microsecond time resolution \cite{Scheibert10_prl,Dalmas13_ijf}: 
\begin{itemize}
\item A hole is drilled at the tip of the seed crack to delay fracture and increase the potential energy stored in the specimen at crack initiation;
\item The time evolution of $v$ is measured by monitoring, via an oscilloscope, the successive rupture of parallel $500\,\mu$m large gold lines deposited on the surface;
\item The time evolution of $K$ is obtained via finite element analysis. That of $G$ is then deduced from Irwin's relation \ref{Sec1:equ7}.
\end{itemize}

\begin{figure}
\begin{center}
\includegraphics[width=0.8\textwidth]{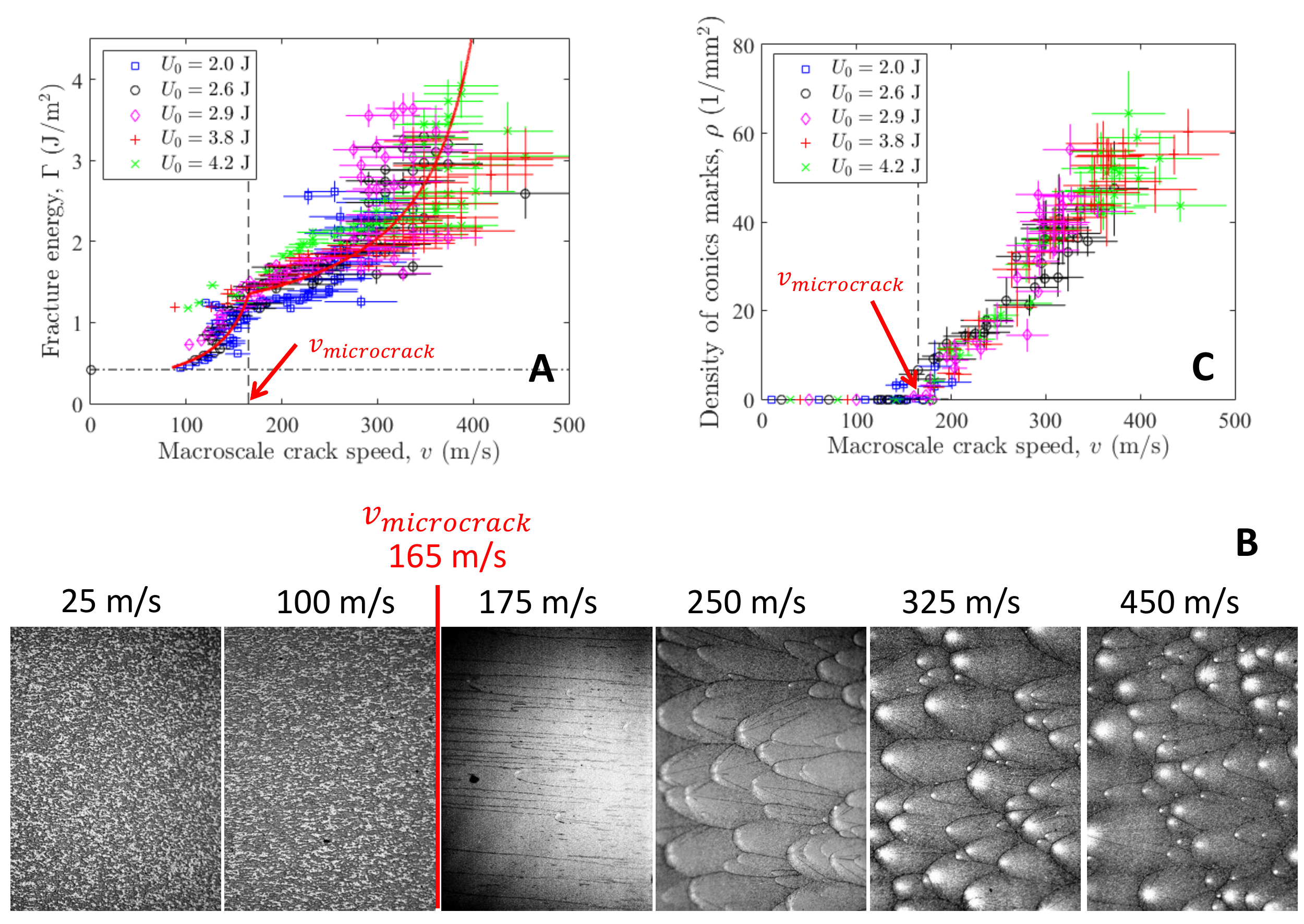}
\caption{Signature of microcracking onset on the dynamic fracture of PMMA. Panel A: Fracture energy $\Gamma$ as a function of the (macroscale) crack speed $v$, for five different experiments at different potential energy at crack initiation, $U_0$. The horizontal dotted line indicates the quasi-static value, $\Gamma(v=0)=420\un{J/m}^2$. The vertical dotted line points out the kink occurring at the microcracking onset, $v_{microcrack} = 165\un{m/s}=0.19 C_R$. Panel B: Sequences of microscope images ($1\times 1.4\un{mm^2}$) showing the evolution of the fracture surfaces as $v$ increases. Beyond $v_{microcrack}$, conics marks are visible and their number increases with $v$. They sign the existence of microcracks forming ahead of the propagating main front. Panel C: Density of conic marks as a function of $v$. In both panels (A) and (C), the vertical dashed line indicates $v_{microcrack}$ and the errorbars indicate a $95\%$ confident interval (adapted from \cite{Scheibert10_prl,Dalmas13_ijf}).}
\label{fig5}
\end{center}
\end{figure}

These series of experiments reveal that, contrary to the LEFM assumptions, $\Gamma$ rapidly increases with $v$ (Figure \ref{fig5}A). The Rayleigh wave speed $c_R$ then stops being the natural limit for $v$, even if equation \ref{Sec1:equ9} is fulfilled. These experiments also reveal the existence of a well-defined critical speed, $v_{microcrack}$, above which the fracture surfaces are decorated with beautiful conical marks (Figure \ref{fig5}B). These conics are known \cite{smekal53_oia,Ravichandar97_jmps} to be  the signature of penny-shape microcracks, growing radially ahead of the main front and subsequently coalescing with it. Their density increases almost linearly with $v - v_{microcrack}$ (Figure \ref{fig5}C). This velocity-driven transition, from a nominally-brittle to a quasi-brittle fracture mode\footnote{Brittle fracture can be of two types: Nominally brittle and driven by the propagation of a single crack or quasi-brittle and involving the formation of multiple microcracks.}, translates into a kink in the $\Gamma\,vs.\,v$ curve at $v=v_{microcrack}$ (Figure \ref{fig5}A). 

The mechanisms underlying the formation of the microcracks at high speeds remain largely unsolved. We used atomic force microscopy (AFM) to search for traces left by these mechanisms on the fracture surfaces. These AFM images revealed the presence of a spherical void of diameter $\sim 200\un{nm}$ right at the initiation point of each microcrack \cite{Guerra09_phd}. This suggests a two-stage process: A spherical cavity first forms at a given point, \eg due to the plastic and/or viscous flow within the highly stressed region of the FPZ. Then, a penny-shape microcrack pops up from it.    

The need to exceed a finite critical speed $v_{microcrack}$ to activate these microcracking mechanisms and the fact that the conics density increases with $v$ are not so easy to interpret. They cannot be explained by postulating the presence of random defects (weak points) which would turn into microcracks when the local stress (or strain) exceeds a given threshold value. Indeed, due to the singular nature of the stress field (equation \ref{Sec1:equ6dyn}), such a criterion would end up being fulfilled for \textit{any} defect encountered by the propagating crack and the conics density would be independent of the crack speed. We hence proposed \cite{Scheibert10_prl} that \textit{two} conditions should be fulfilled to make a microcrack pop-up:
\begin{itemize}
\item The stress at the considered defect is larger than a threshold value;
\item The considered defect is sufficiently far from the main crack front to allow sufficient time for the nucleated microcrack to reach maturity. 
\end{itemize}
Then, $K$ should be larger than a finite threshold value to activate microcracking (see equation Sec1:equ6dyn), which imposes a finite threshold velocity $v_{microcrack}$. It also makes the conics density increase with $K$, and hence with $v$. The rationalization of the above conditions was found \cite{Scheibert10_prl} to reproduce the $\rho\,vs.\,v$ curve shown in figure \ref{fig5}C fairly well.  

As a final remark to this section, let us point out the fact that microcracks forming ahead of a dynamically growing crack have been evidenced in a variety of materials: In most brittle polymers \cite{Ravichandar98_ijf,Du10_jms}, in rocks \cite{Ahrens93_jgr}, in oxide glasses \cite{Rountree10_pcge} in some nanophase ceramics and nanocomposite \cite{Rountree02_armr}, in metallic glasses \cite{Murali11_prl}, etc. This yields us to argue this switch from nominally-brittle to quasi-brittle mode at high speed is a generic mechanism in the dynamic fracture of disordered solids.

\subsection{From local front velocity to apparent macroscopic speed of cracks}\label{Sec3.2}

\begin{figure}
\begin{center}
\includegraphics[width=\textwidth]{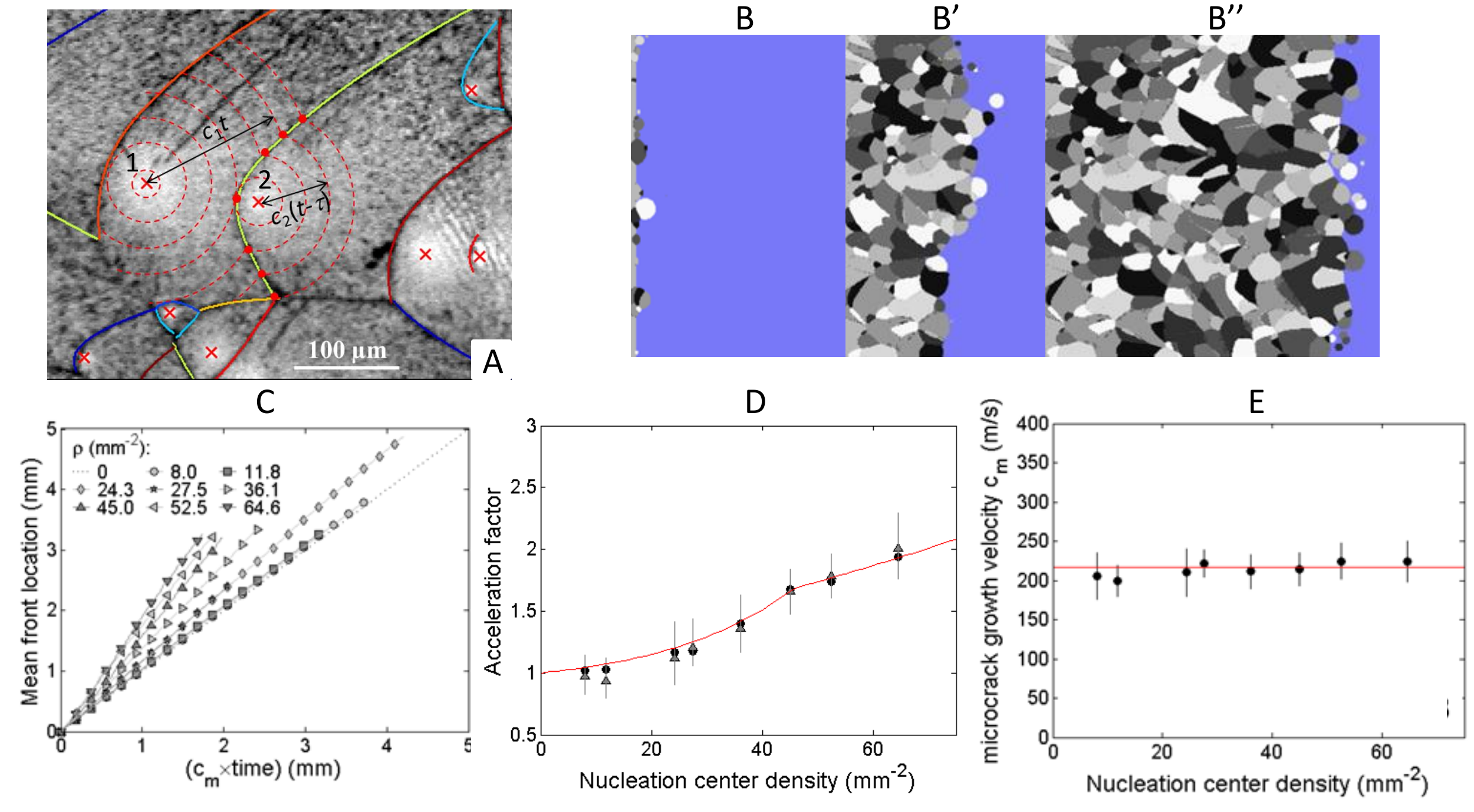}
\caption{From local front speed within FPZ to apparent speed at the continuum scale. Data are for PMMA. Panel A: Reconstruction scheme of the microscale damage dynamics from the post-mortem fracture surfaces. The bright white regions provide the nucleation centers (red $\times$). Red dots sketch the successive positions of two growing microcracks, denoted by (1) and (2). The crossing points give rise to the green branch of the conic mark. The fit of this branch permits to infer both the ratio $c_2/c_1$ of the microcrack speeds and the time interval $t_2-t_1$ between the two nucleation events. From the nucleation positions, the speed ratio $c_j/c_i$ and the inter-nucleation times $t_j-t_i$, it is possible to reconstruct the time-space dynamics of microcracking events, within nanosecond and micrometer resolution. Panels B$\rightarrow$B'' show such a reconstructed sequence. The blue part is the uncracked material and the grey one is the cracked part. The different gray levels illustrate the fact that the fracture surface does not result from the propagation of the main crack front, but is the sum of the surfaces created by each microcrack. A different gray level has been randomly assigned to each of them. The analysis of these reconstructions has shown that all microcracks grow with the same velocity $c_m$. Panel C: Evolution of the mean crack front as a function of $c_m \times t$ for different microcrack density. The slope of these curves provides the ratio between the apparent macroscale crack speed $v$ and the true local speed $c_m$ of the propagating (micro)crack front. This boosting factor is plotted as a function of microcrack density in panel D. Panel E: Deduced variation of $c_m$ with $\rho$. Horizontal red line indicated the mean value $c_m=217\un{m/s}=0.24 c_R$ (Adapted from \cite{Guerra12_pnas,Dalmas13_ijf}).}
\label{fig6}
\end{center}
\end{figure}

The careful analysis of the PMMA fracture surfaces has also permitted us to uncover the selection of fracture speed in this quasi-brittle regime \cite{Guerra12_pnas}. Indeed, from the conics pattern, it has been possible to determine the nucleation center of each microcrack, the time at which they nucleated and the speed at which they grew (Figure \ref{fig6}A and \cite{Dalmas13_ijf} for details). This allowed us to reconstruct the complete spatiotemporal microcracking dynamics underlying fast fracture, with micrometer/nanosecond resolution (Figure \ref{fig6}B$\rightarrow $B''). These reconstructions demonstrate that the main front does not progress regularly, but by successive jumps of finite length. Those correspond to the coalescence events of the main front with a microcrack growing ahead. This mechanism makes the effective velocity measured at the macroscopic scale, $v$, larger than the "true" local propagating speed $c_m$ of the individual (micro)crack fronts. The ratio between the two increases with the microcrack density, $\rho$ (Figures \ref{fig6}C and D). From the knowledge of $v$ and $\rho$, it has been possible \cite{Guerra12_pnas} to infer the value of $c_m$ in all our experiments and, surprisingly, it has been found to be constant and equal to a fairly low value: $c_m=210\un{m/s} = 0.24 c_R$ (Figure \ref{fig6}E). In other words, the fairly large velocities observed above $v_{microcrack}$ in PMMA (Figure \ref{fig5}A) are not to be attributed to that of a crack line described by LEFM and equation \ref{Sec1:equ9}. They result from a collective effect yielded by the coalescence of microcracks with each other within the FPZ. This boost mechanism demonstrated here on PMMA likely arises in all the situations involving propagation-triggered microcracks. It may also be at play in the intersonic shear rupture and earthquakes \cite{Tarasov08_ijrmms}. It is finally worth to note that a similar mechanism has been recently reported on simulations of ductile fracture \cite{Osovski15_jmps}.

The analysis of the $\Gamma\,vs\,v$ curve in the low speed regime ($v\leq v_{microcrack}$ in figure \ref{fig5}A) has permitted to uncover the origin of the limiting propagating speed $c_m$ in PMMA \cite{Dalmas13_ijf}. Once recasted into a $\Gamma\,vs\,G$ curve, it reveals that $\Gamma$ is proportional to $G$, which is also proportional to the FPZ size\footnote{Calling $\xi$ the process zone size and $\sigma_Y$ the yield stress of the considered material (the stress above which the rheology stops being elastic and the dissipative processes are activated), equation\ref{Sec1:equ8} predicts $\xi=K^2/2\pi \sigma_Y^2$. Using Irwin's relation \ref{Sec1:equ7}, one gets $\xi=E G/2\pi \sigma_Y^2$.}. As a consequence, it is not the energy dissipated \textit{per unit surface} which is constant here, but the energy dissipated \textit{per unit volume of FPZ}. The rationalization of this statement has provided \cite{Dalmas13_ijf} a relation $\Gamma(c)$ ($c$ here refers to the true local speed of individual (micro)crack fronts). This relation yields a divergence of $\Gamma$ at a finite value which was related \cite{Dalmas13_ijf} to some of the material constants, namely the dilatational and Rayleigh wave speed, the Young's modulus, the energy dissipated per unit volume within the FPZ, and the yield stress: Numerical application of the so-obtained formula gives $c_m=204\un{m/s}$ in PMMA.  

\subsection{Microbranching instabilities at high speed}\label{Sec3.3}

Another element of complexity arises at even higher speed: In many materials including PMMA, the crack front splits into a succession of secondary cracks known as microbranches when the crack speed $v$ gets larger than a second critical velocity $v_{microbranch} \sim 0.4 c_R$ \cite{Fineberg91_prl,Fineberg99_pr}. These secondary cracks are short-lived; they rapidly stop and remain confined along the main crack. Furthermore, they do not extend over the entire thickness of the specimen but are spatially localized. This microbranching instability has two consequences of importance: It leads to rough fracture surfaces and $\Gamma$ stops being a function of $v$ only \cite{Sharon99_nature}. 

Careful measurements \cite{Sharon99_nature,Livne05_prl} of the normalized value $v_{microbranch}/c_R$ reveal that:
\begin{itemize}
\item It slightly depends on the considered material (e.g. $0.36$ in PMMA and $0.42$ in oxide glasses \cite{Sharon99_nature}); 
\item In polyacrylamide gels, it increases (roughly linearly) with the crack acceleration $\dot{v}$ \cite{Livne05_prl};
\item In polyacrylamide gels, it decreases with the specimen thickness \cite{Livne05_prl}. 
\end{itemize}
These observations underly \cite{Livne05_prl} the presence of an activation mechanism: For crack speed above a critical value $0.4 c_R$, the random perturbations intrinsic to the system give rise to a microbranch with a finite probability; this activation is all the more likely as the time left to the process is large (i.e. $\dot{v}$ is small) and the number of potential activation sites is large (i.e. the thickness of the specimen is large). The experiments reported in the previous section have also shown \cite{Guerra12_pnas} that in PMMA, $v_{microbranch}$ coincides with the moment when the microcrack density is large enough so that the microcracks can no longer pop up one by one, but are formed by cascades. Very recent experiments in gels \cite{Boue15_prl} also suggest to relate this instability to an oscillatory instability observed at a much higher speed \cite{Livne08_prl}, itself related to the non-linear elasticity of the gel \cite{Bouchbinder08_prl}. In a nutshell, despite significant advances on this problem, the origin of this microbranching instability remains largely unsolved.    

\section{Conclusion and open challenges}\label{concl}

Stress enhancement at crack tips and so-induced sensibility to microscale defects  make the problem of brittle fracture difficult to tackle. In this article, we first briefly reviewed the strategy implemented by the standard continuum fracture theory -- linear elastic fracture mechanics (LEFM) -- to bypass the problem. By reducing the question of how a solid breaks to that of how a preexisting crack propagates into the solid, it provides a coherent and powerful framework based on linear elasticity to describe when and how fast cracking occurs in a quantitative and predictable manner. Still and despite its success, LEFM fails in explaining the highly intermittent dynamics sometimes observed in the slow fracturing regime of heterogeneous materials. It also falls short in capturing the anomalously high speeds observed in the dynamic fracture of amorphous materials.

The latter has been studied via dedicated experiments in PMMA, the specificity of which have been to give access to both the continuum-level scale dynamics (via experimental mechanics tools) and the microscale one (via fractographic reconstruction). Beyond a given velocity, the propagation of the crack is accompanied by a multitude of microcracks forming ahead of the main front. And the coalescence of these microcracks with the main front boosts the apparent fracture speed when measured at the macroscopic scale! As briefly discussed in section \ref{Sec3.1}, the mechanisms underpinning the  formation of these microcracks are only partially understood. A major difficulty here is that LEFM allows describing the growth of a preexisting crack, but not the initiation of a crack or a microcrack. The fast development of \textit{Finite Fracture Mechanics (FFM)} initiated by D. Leguillon at the beginning of the 2000's \cite{Leguillon01_cras,Leguillon02_ejms} appears promising in this field. Indeed, this framework extends the classical approach of fracture mechanics, and makes it able to tackle the problem of crack initiation by completing the Griffith's energy based criterion for fracture with a second criterion comparing the tensile stress with the tensile strength. FFM may offer the proper tools to understand and subsequently model the conditions for microcrack formation in this dynamic fracture regime. 

Back to the crackling dynamics observed in slow fracture, the past ten years have seen the emergence of concepts from non-linear physics which, combined with continuum fracture mechanics and elasticity framework, offer a promising framework to address the problem: The depinning interface approach, presented in section 3.2, has succeeded to capture, at least qualitatively, the statistics of the dynamics fluctuations. It also provides rationalized tools to predict when crackling will occur. Note that the agreement between theory and experimental observations remains qualitative only and the exponent values, in particular, are different. This is likely due to the fact that current depinning approaches consider specimens of infinite thickness. The availability of kernels (term $J$ in equation \ref{Sec2:equ3}) taking into account explicitly the specimen thickness is currently missing.  

As briefly discussed in section section \ref{Sec3.3}, microbranching instabilities develop at high speeds and make the crack dynamics highly fluctuating also in the dynamic fracture regime; then, it cannot be described via LEFM anymore. A promising avenue to bypass the problem is to develop stochastic equations of motion based on interface growth models, along the lines applied successfully in the slow fracture regime. A major bottleneck here is to capture correctly the dynamic stress transfers through acoustic waves, occurring as a dynamically growing crack interacts with the material disorder \cite{Boudet98_prl,Sharon01_nature,Bonamy03_prl,Bonamy05_ijf}. The availability of analytical solutions for weakly distorted cracks within the full 3D elastodynamics framework \cite{Willis95_jmps,Willis97_jmps} suggests promising developments in the not-too-distant future (see \cite{Ramanathan97b_prl,Bouchaud02_jmps,AddaBedia13_prl} for past theoretical attempts in this context).  

In the form presented in sections \ref{Sec2.2} and \ref{Sec2.3}, the depinning approach of fracture suffers from several limitations that, finally, deserve to be commented. First, it incorporates the effect of material inhomogeneities only on the fracture properties; the role played by a contrast in term of Young's modulus, for instance, is ignored. This prevents the approach to be applied to composite materials, for instance, when hard particles or fibers are embedded in a softer matrix. Second, it was derived within the {\em isotropic} linear elasticity framework and, as such, cannot describe laminar materials. Third, it presupposes the existence of a well-defined upper limit for the size of the material inhomogeneities, much smaller than the macroscopic dimensions involved in the problem. As such, the depinning framework cannot address the complex hierarchical structures encountered in biological systems or bio-inspired ones, such as bones for instance. The proof of concept and the promises offered by these statistical approaches of fracture are now established. A formidable challenge for the future will be to overcome the above limitations -- and most likely others to discover --, and then make these approaches based on statistical physics applicable to the high-performance materials and complex structures of engineering and technological interest.

\section*{Acknowledgements}

The author is particularly indebted to Jonathan Barés and Claudia Guerra: Many of the results presented here were obtained during their PhD thesis. He also wish to thank his colleagues, postdocs and students for their contributions, namely Luc Barbier, Fabrice Célarié, Davy Dalmas, Alizée Dubois, Lamine Hattali, Véronique Lazarus, Arnaud Lesaine, Laurent Ponson, Cindy Rountree and Julien Scheibert. He is also grateful to Jacques Villain and Cindy Rountree for the critical reading of this review and their many suggestions to improve the didacticism. Support from various funding agencies is also acknowledged: Agence Nationale de la Recherche (Grants RUDYMAT No. ANR-05-JCJC-0088 and MEPHYSTAR No. ANR-09-SYSC-006-01), the RTRA Triangle de la Physique (Grant CODERUP No 2007-46) and the "Investissement d'Avenir" LabEx PALM (Grant Turb$\&$Crack No. ANR-10-LABX-0039-PALM).


\end{document}